\documentclass[12pt]{iopart}
\usepackage{graphicx}% Include figure files

\usepackage{color}

\usepackage{iopams}

\begin{document}

\title[Imaging magnetic scalar potentials]{Imaging magnetic scalar potentials by laser-induced fluorescence from bright and dark atoms}

\author{I Fescenko$^{1,2}$ and A Weis$^1$}

\address{ $^{1}$ Physics Department, University of Fribourg, Chemin du Mus\'ee 3, 1700 Fribourg, Switzerland}
\address{ $^{2}$ Institute of Atomic Physics and Spectroscopy, University of Latvia, 19 Rainis Boulevard, LV-1586 Riga, Latvia\\}
\ead{iliafes@gmail.com}

\date{\today}

\begin{abstract}
We present a spectroscopic method for mapping two-dimensional
distributions of magnetic field strengths (magnetic scalar potential
lines) using CCD recordings of the fluorescence patterns emitted by
spin-polarized Cs vapor in a buffer gas exposed to
inhomogeneous magnetic fields.
The method relies on the position-selective destruction of
spin polarization by magnetic resonances induced by multi-component
oscillating magnetic fields, such that magnetic potential lines can
directly be detected by the CCD camera.
We also present a generic algebraic model allowing the calculation
of the fluorescence patterns and find excellent agreement with the
experimental observations for three distinct inhomogeneous field
topologies.
The spatial resolution obtained with these
proof-of-principle experiments is on the order of 1~mm.
A substantial increase of spatial and magnetic field resolution is expected by
deploying the method in a magnetically shielded environment.

\end{abstract}

%33.57.+c Magneto-optical and electro-optical spectra and effects
%33.80.Be Level crossing and optical pumping
%\keywords{magneto-optical effects,magnetic field imaging,cesium}

%\submitto{\JPD}
\maketitle

\section{\label{sec:Introduction}Introduction}
The use of inhomogeneous magnetic fields for encoding inhomogeneous distributions of spin-polarized particles into a (Larmor) frequency spectrum is the underlying principle of magnetic resonance imaging (MRI).
In this paper we apply the converse of this procedure by using a homogeneous distribution of polarized particles for imaging inhomogeneous magnetic field distributions.
In the reported proof-of-principle experiments we demonstrate that the method yields high quality images of magnetic potential lines produced by known sources.
While conventional medical MRI relies on detecting the relevant Larmor
frequency spectrum by pick-up coils, the imaging sensor described here uses a spin-polarized alkali atom vapor in combination with optical detection.

Our research is motivated by two main objectives.
On one hand we wanted to develop a simple lecture demonstration experiment for visualizing dark atomic states, hence the fact that the experiment were carried out in a magnetically unshielded environment offering visibility of all experimental components.
On the other hand, we are currently developing means for imaging magnetic nanoparticles embedded in biological tissues in view of biomedical applications.
While the proof-of-principle experiments reported below fully satisfy the first objective, they constitute a first step towards the second objective that aims at imaging inhomogeneous nanoparticle distributions  via the inhomogeneous magnetic field patterns that they produce.
A magnetic field imaging camera based on this principle will open new  pathways for quantitative space- and time- resolved studies of the distribution and dynamics of nanoparticles in biological samples.

\section{\label{sec:Method}Method}
It is well known that the optical absorption coefficient of an atomic ensemble depends on the degree and nature of its spin polarization.
In general, the creation of spin polarization by optical pumping with (resonant) polarized light reduces the absorption coefficient of an atomic medium, which thus becomes more transparent for the pumping light.
This effect is a manifestation of electromagnetically-induced transparency (EIT) that goes in pair with a reduction of atomic fluorescence, a fact which is commonly expressed by stating that the polarized medium is in a dark state.
Dark atomic ensembles are also said to be in a coherent population trapped (CPT) state since optical pumping creates a coherent superposition of (Zeeman or hyperfine) magnetic sublevels, with amplitudes and phases such that the incident polarized light does not couple to that particular superposition state.
The seminal paper by Alzetta \textit{et al.}~\cite{Alzetta1976} nicely illustrated the concept of (non-absorbing) dark states by photographic recordings of their reduced fluorescence.

Imaging inhomogeneous distributions of spin-polarized atoms have been used in the past for space-time resolved studies of atomic diffusion \cite{Skalla1997,Ishikawa1999,Giel2000}.
Magnetic iso-field lines were mapped for the first time by Tam and Happer \cite{Tam1977} by photographic recordings of fluorescene from Na vapor excited by a multi-frequency optical field (frequency comb) in homogeneous and inhomogeneous fields.
They applied the technique for the visual analysis of optical \textit{rf} spectra \cite{Tam1977,Tam1979}.
More recently quantitative digital methods based on CCD cameras were deployed for field imaging in fluorescence \cite{Asahi2003} and transmission~\cite{Mikhailov2009} experiments.
In contrast to previous field mapping experiments that used bi- or poly-chromatic optical fields to prepare non-absorbing (non-fluorescing) states our experiments use monochromatic laser radiation.
Optical pumping with \textit{linearly polarized} light on the $F_g{=}4{\rightarrow}F_e{=}3$ component of the Cs $D_1$ transition creates spin alignment oriented along the direction of light polarization, while pumping with \textit{circularly polarized} light prepares both vector polarization (orientation) and tensor polarization (alignment) along the light propagation direction $\hat{k}$, the alignment contribution being negligible compared to the orientation on that specific hyperfine transition \cite{Grujic2013}.
A static magnetic field applied to a polarized medium affects the degree and orientation of the medium's polarization, an effect known as ground state Hanle effect (GSHE).
Since the fluorescence intensity depends both on the magnetic field amplitude and its relative orientation with respect to the direction of spin polarization, an inhomogeneous magnetic field therefore 
produces a spatially varying fluorescence pattern (Hanle background pattern).
Irradiating the sample in the inhomogeneous magnetic field additionally with a
mono- or poly-chromatic \textit{rf} field will depolarize the sample
at spatial locations where the local Larmor frequency matches the
\textit{rf} frequency, thereby lighting up the regions with a
specific modulus of the magnetic field strength.
In this way, the recording of the fluorescence patterns by a
CCD camera allows the mapping of magnetic potential lines.
%

%%%%%%%%%%%%%%%%%%%%%%%%%%%%%%%%%%%%%%%%%%%%%%%%%%%%%%%%%%%%
\section{Laser-induced fluorescence of spin-polarized atoms }
%%%%%%%%%%%%%%%%%%%%%%%%%%%%%%%%%%%%%%%%%%%%%%%%%%%%%%%%%%%%
\label{sec:CircPolFluo}
The power of fluorescence light induced by a resonant laser of power
$P_0$ traversing a \textit{spin-oriented} atomic vapor is given by
\begin{equation}
dP_{f}^{(1)}=P_0\,\kappa_0
L\,(1-\alpha^{(1)}\,m_{1,0})\,f^{(1)}(\Omega)\,d\Omega\,
\end{equation}
where $\kappa_0$ is the peak optical absorption coefficient of the
(in our case, Doppler-broadened) atomic absorption line, $L$ the
length of the illuminated vapor column imaged onto the fluorescence
detector, and $d\Omega$ the solid angle intercepted by the detector.
The superscript $(1)$ refers to $k{=}1$ and denotes the vector
character of the spin polarization (orientation).
The function $f^{(1)}(\Omega)$ represents the angular distribution of the fluorescence intensity.
Throughout the paper we will assume an isotropic distribution of
fluorescence, thus choosing  $f^{(1)}(\Omega)$ to be a constant.
The parameter $\alpha^{(1)}=\alpha_{F_g,F_e}^{(1)}$ is the
orientation analyzing power introduced in Ref.~\cite{Castagna2011},
which depends on the angular momentum quantum numbers of the
specific transition excited by the laser.
We characterize spin orientation in terms of the longitudinal vector
multipole moment $m_{1,0}$, the only anisotropy parameter of the
medium to which circularly polarized light couples when the medium's
alignment can be neglected.

With the above we can express the total fluorescence power recorded by the detector as
%
%\%begin{equation}
%P_{f}^{(1)}=\int\frac{dP_{f}^{(1)}}{d\Omega}\,d\Omega\equiv
%A^{(1)}-B^{(1)}\,m_{1,0}\,, \label{eq:Pf10}
%\end{equation}
%%
%
\begin{equation}
P_{f}^{(1)}=\int\frac{dP_{f}^{(1)}}{d\Omega}\,d\Omega\equiv
A^{(1)}\left(1-\alpha^{(1)}\,m_{1,0}\right)\,. \label{eq:Pf10}
\end{equation}

In the same way, one can represent the fluorescence of spin-aligned
atoms by
\begin{equation}
P_{f}^{(2)}=\int\frac{dP_{f}^{(2)}}{d\Omega}\,d\Omega\equiv
A^{(2)}\left(1-\alpha^{(2)}\,m_{2,0}\right)\,, \label{eq:Pf20}
\end{equation}
where the atomic multipole moment $m_{2,0}$ is the longitudinal spin
alignment that is oriented along the light polarization when
produced by optical pumping with linearly polarized light.
$\alpha^{(2)}$ is the alignment analyzing power, introduced in
\cite{Breschi2012}.

%%%%%%%%%%%%%%%%%%%%%%%%%%%%%%%%%%%%%%%%%%%%%%%%%%%%%%%%%%%%
\subsection{\label{sec:HanleCircPol}Fluorescence of spin-oriented atoms in an
inhomogeneous magnetic field}
%%%%%%%%%%%%%%%%%%%%%%%%%%%%%%%%%%%%%%%%%%%%%%%%%%%%%%%%%%%%
%
 A magnetic field of arbitrary
direction applied to the medium will change the magnitude and
orientation of its spin polarization.
It was shown in Ref.~\cite{Castagna2011} that the steady-state value
$m_{1,0}^{ss}$ that the longitudinal orientation reaches by virtue
of the GSHE is given by the Hanle function
\begin{equation}
\mu^{(1)}(\beta_\parallel,\beta_\perp)\equiv\frac{m_{1,0}^{ss}}{m_{1,0}^{eq}}
=\frac{1+\beta_\parallel^2}{1+\beta_\parallel^2+\beta_\perp^2}\,,
\label{eq:m10Hanle}
\end{equation}
where $\beta_\parallel$ and $\beta_\perp$ denote the longitudinal and transverse (with respect to $\vec{k}$) magnetic field components, expressed in dimensionless units according to
\begin{equation}
\beta_\parallel\equiv\frac{\omega_\parallel}{\gamma}=\frac{\gamma_{F}\,|\vec{B}_\parallel|}{\gamma}
\quad\rm{and}\quad
\beta_\perp\equiv\frac{\omega_\perp}{\gamma}=\frac{\gamma_{F}\,|\vec{B}_\perp|}{\gamma}\,,\label{eq:betraDef}
\end{equation}
where we have assumed identical longitudinal and transverse
orientation relaxation rates ($\gamma_1=\gamma_2\equiv\gamma$).
The Larmor frequencies $\omega_{\parallel,\perp}$ are related to the respective field components via the gyromagnetic ratio $\gamma_{F}\approx (2\pi)$~3.5~Hz/nT of the Cs ground state.
In Eq.~\eref{eq:m10Hanle}, $m_{1,0}^{eq}$ represents the
longitudinal orientation that is achieved by optical
pumping in a polarization stabilizing longitudinal field
($\beta_\parallel\gg\beta_\perp$), so that
$\mu^{(1)}(\beta_\parallel,\beta_\perp)$ can assume values between
$0$ and $1$.
%
%We note that when dealing with orientation created (and probed) by
%circularly polarized light, $\beta_\parallel$ refers to the magnetic
%field component along the laser propagation direction $\hat{k}$,
%while $\beta_\perp$ is the modulus of the field perpendicular to
%$\vec{k}$.

When the atomic medium is exposed to an inhomogeneous magnetic
field, $\beta_\parallel$ and $\beta_\perp$ depend on the position
$\vec{r}$ in the atomic medium from which fluorescence is emitted,
so that the florescence power emitted by a volume element (voxel) of
the atomic medium located at $\vec{r}$ can be written as
\numparts
\begin{eqnarray}\label{eq:Pf10All}
P_{f}^{(1)}(\vec{r})
&=&A^{(1)}\left(1-\alpha^{(1)}\,m_{1,0}^{eq}\,\mu^{(1)}(\beta_\parallel,\beta_\perp)\right)\label{eq:Pf10A}\\
&\propto &1-C^{(1)}\,\frac{1+\beta_\parallel^2(\vec{r})}{1+\beta_\parallel^2(\vec{r})+\beta_\perp^2(\vec{r})}
\,,\label{eq:Pf10B}
\end{eqnarray}
\endnumparts

with
\begin{equation}
C^{(1)}=\frac{B^{(1)}}{A^{(1)}}=\frac{\alpha^{(1)}\,m_{1,0}^{eq}}
{P_0\,\kappa_0L\,\int{f^{(1)}(\Omega)d\Omega}}\,.
 \label{eq:C1}
\end{equation}

In the experiments reported below, we use a CCD camera for recording
the fluorescence emitted from a quasi two-dimensional volume excited
by a sheet of laser light in a cubic vapor cell exposed to an
inhomogeneous magnetic field.
The light intensity distribution on the CCD chip can then be
calculated by inserting the (known) spatial dependence of
$\beta_\parallel(\vec{r})$ and $\beta_\perp(\vec{r})$ in the light
emitting volume (located at position $\vec{r}$) into
Eq.~\eref{eq:Pf10B} .

%%%%%%%%%%%%%%%%%%%%%%%%%%%%%%%%%%%%%%%%%%%%%%%%%%%%%%%%%%%%%%%%%%%%%%%%%%%%%%%%%%%%%%
\subsection{\label{sec:rfDepol}Magnetic resonance induced modifications of the fluorescence from oriented atoms}
%%%%%%%%%%%%%%%%%%%%%%%%%%%%%%%%%%%%%%%%%%%%%%%%%%%%%%%%%%%%%%%%%%%%%%%%%%%%%%%%%%%%%%
%
Polarized atoms emit a weak fluorescence, while unpolarized atoms
emit a stronger fluorescence.
Local magnetic fields that are oriented along the axis of spin
polarization stabilize the latter, so that voxels containing
polarized atoms lead to darker parts in the  CCD image.
The experiments described below show that the models developed here
yield a good quantitative prediction of the observed fluorescence
from atoms in a variety of inhomogeneous fields.
However, the inverse problem which consists in inferring the
magnetic field distribution from the recorded fluorescence pattern
is a more demanding task.
Although this inverse problem may, in principle, be solved based on
fluorescence images recorded with different (circular and linear)
polarizations, we have not yet attempted to do so.

Rather than attempting to solve the inverse problem we present here a method allowing the easy visualization and
measurement of magnetic potential lines.
The method is based
on the space-selective destruction of spin polarization by magnetic
resonance.
It is well known that a weak oscillating magnetic field
$B_1\cos(\omega_{rf}t)$, called `$rf$'-field, resonantly modifies
the magnetization (spin polarization) of a medium, when its
oscillation frequency matches the Larmor frequency, i.e., when
$\omega_{rf}{=}\omega_L(\vec{r})$,  a phenomenon known as magnetic resonance.
The steady-state solution of the Bloch equations (under the
assumption of identical longitudinal and transverse relaxation
rates) is given by the well-known expression~\cite{Bloch1946}
\begin{equation}
m_{1,0}=m_{1,0}^{ss}\,\left(1-\frac{\sqrt{S_{rf}}}{\delta^2+S_{rf}+1}\right)
 \label{eq:rfm10}
\end{equation}
for the longitudinal spin orientation, with
$\delta=(\omega_{rf}-\omega_L)/\gamma\equiv\beta_{rf}-\beta$, where
$\gamma$ is the polarization's relaxation rate.
The \textit{rf} saturation parameter $S_{rf}$ is defined as
\begin{equation}
S_{rf}=\left(\frac{\gamma_F\,B_1}{\gamma}\right)^2\equiv\left(\frac{\beta_1}{\gamma}\right)^2\,.
\end{equation}
In Eq.~\eref{eq:rfm10}, the multipole moment $m_{1,0}^{ss}$ represents the steady-state
spin polarization established by the Hanle effect in absence of the $rf$ field ($S_{rf}{=}0$), given by Eq.~\eref{eq:m10Hanle}.
The orientation resulting from the joint depolarization by the
inhomogeneous field and the rf field is thus given by
\begin{equation}
m_{1,0}=m_{1,0}^{eq}\,\frac{1+\beta_\parallel^2}{1+\beta_\parallel^2+\beta_\perp^2}\,
\left(1-\frac{\sqrt{S_{rf}}}{\delta^2+S_{rf}+1}\right)\,,
 \label{eq:rfm102}
\end{equation}
\pagebreak
and the fluorescence of equation~\eref{eq:Pf10A} becomes
\numparts

\begin{eqnarray}\label{eq:Pf10Final}
P_{f}^{(1)}(\vec{r})
&=&A^{(1)}\left(1-\alpha^{(1)}\,m_{1,0}\right)\label{eq:Pf10Final1}\\
&=&A^{(1)}\left[1-\alpha^{(1)}\,m_{1,0}^{eq}\,\frac{1+\beta_\parallel^2}{1+\beta_\parallel^2+\beta_\perp^2}\,
\left(1-\frac{\sqrt{S_{rf}}}{\delta^2+S_{rf}+1}\right)\right]\label{eq:Pf10Final2}\\
&=&A^{(1)}-B^{(1)}\,\frac{1+\beta_\parallel^2}{1+\beta_\parallel^2+\beta_\perp^2}\,
\left(1-\frac{\sqrt{S_{rf}}}{\delta^2+S_{rf}+1}\right)\label{eq:Pf10Final3}\\
&\propto &1-C^{(1)}\,\frac{1+\beta_\parallel^2}{1+\beta_\parallel^2+\beta_\perp^2}\,
\left(1-\frac{\sqrt{S_{rf}}}{(\beta_{rf}-\beta)^2+S_{rf}+1}\right)\label{eq:Pf10Final4}\,,
\end{eqnarray}
\endnumparts
which reduces to equation~\eref{eq:Pf10B} for $S_{rf}=0$.

In an inhomogeneous field $\vec{\beta}(\vec{r}){=}\vec{\beta}_\parallel(\vec{r})+\vec{\beta}_\perp(\vec{r})$, the parameters $\beta_\parallel$, $\beta_\perp$,  $\beta{=}\sqrt{\beta_\parallel^2+\beta_\perp^2}$ in Eq.~\eref{eq:Pf10Final4} depend on $\vec{r}$, while $S_{rf}\propto\beta_1^2$ is assumed assumed to be homogeneous.
When the spin-polarized atomic medium in an inhomogeneous magnetic
environment is irradiated with an $rf$ field, this field will
depolarize the medium at positions $\vec{r}$ at which the modulus $\beta$ of
the local magnetic field field $\vec{\beta}$ obeys
$\beta=\beta_{rf}=\omega_{rf}/\gamma_F$, so that these regions will light up
with an enhanced intensity in the CCD image.
When light from a plane in the vapor is imaged onto the CCD chip, it
is possible in this way to visualize the lines of constant
$|\vec{\beta}|$ in that plane.

Note that the lines of constant $|\vec{\beta}|$ represent the the
magnetic scalar potential $\Psi(\vec{r})$, from which the magnetic
field (in current-free regions) can be derived via
$\vec{B}(\vec{r})=-\mu_0\,\vec{\nabla}\Psi(\vec{r})$.
Note also that throughout the paper we refer to the magnetic
induction vector $\vec{B}$ as `magnetic field' vector, following
common laboratory practice.
%
%%%%%%%%%%%%%%%%%%%%%%%%%%%%%%%%%%%%%%%%%%%%%%%%%%%%%%%%%%%%
\subsection{Magnetic resonance induced fluorescence of spin-aligned atoms in an
inhomogeneous magnetic field}
\label{sec:LinPolFluo}
%%%%%%%%%%%%%%%%%%%%%%%%%%%%%%%%%%%%%%%%%%%%%%%%%%%%%%%%%%%%
%
Optical pumping with linearly polarized light produces a longitudinal alignment $m_{2,0}^{eq}$ that is oriented along the light polarization.
In Ref.~\cite{Breschi2012} it was shown that a magnetic field of
arbitrary magnitude and orientation yields a steady-state
magnetization $m_{2,0}^{ss}$ given by
\begin{eqnarray}
\mu^{(2)}(\beta_\parallel,\beta_\perp)\equiv\frac{m_{2,0}^{ss}}{m_{2,0}^{eq}}
=\frac{1}{4}+
\frac{3}{4}\,\frac{1+8\beta_\parallel^2+16\beta_\parallel^4}{1+4\beta_\parallel^2+4\beta_\perp^2}
-3\,\frac{\beta_\parallel^2+\beta_\parallel^4}{1+\beta_\parallel^2+\beta_\perp^2}\,,
\label{eq:m20Hanle}
\end{eqnarray}
where $m_{2,0}^{eq}$ represents the longitudinal alignment, oriented
along the laser polarization that is produced by optical pumping in
a stabilizing ($\beta_\parallel{\gg}\beta_\perp$) field, and where
$0{\le}\mu^{(2)}{\le}1$.
Note that in the case of linearly polarized pump and probe light, $\beta_\parallel$ refers to
the magnetic field component along the laser polarization, while
$\beta_\perp$ is the modulus of the field perpendicular to the
polarization.
In analogy with the derivation above, magnetic resonance transitions will then yield a fluorescence signal given by
%
%\begin{widetext}

\begin{eqnarray}
\fl P_{f}^{(2)}(\vec{r})
\propto 1&-&C^{(2)}\,
\left(\frac{1}{4}+
\frac{3}{4}\,\frac{1+8\beta_\parallel^2+16\beta_\parallel^4}{1+4\beta_\parallel^2+4\beta_\perp^2}-
3\,\frac{\beta_\parallel^2+\beta_\parallel^4}{1+\beta_\parallel^2+\beta_\perp^2}\right)\times\nonumber\\
&\times& \left(1-\frac{\sqrt{S_{rf}}}{(\beta_{rf}-\beta)^2+S_{rf}+1}\right),\,
\label{eq:Pf20Final}
\end{eqnarray}
%\end{widetext}
%
with
\begin{equation}
C^{(2)}=\frac{\alpha^{(2)}\,m_{2,0}^{eq}}
{P_0\,\kappa_0L\,\int{f^{(2)}(\Omega)d\Omega}}\,.
 \label{eq:C2}
\end{equation}
%

%%%%%%%%%%%%%%%%%%%%%%%%%%%%%%%%%%%%%%%%%%%%%%%%%%%%%%%%%%%%%%%%%%%%%%%%%%%%%%%%%%%%%
\section{Experimental apparatus}
\label{sec:setup}
\begin{figure*}[t]
    \centering
       \resizebox{\textwidth}{!}{\includegraphics{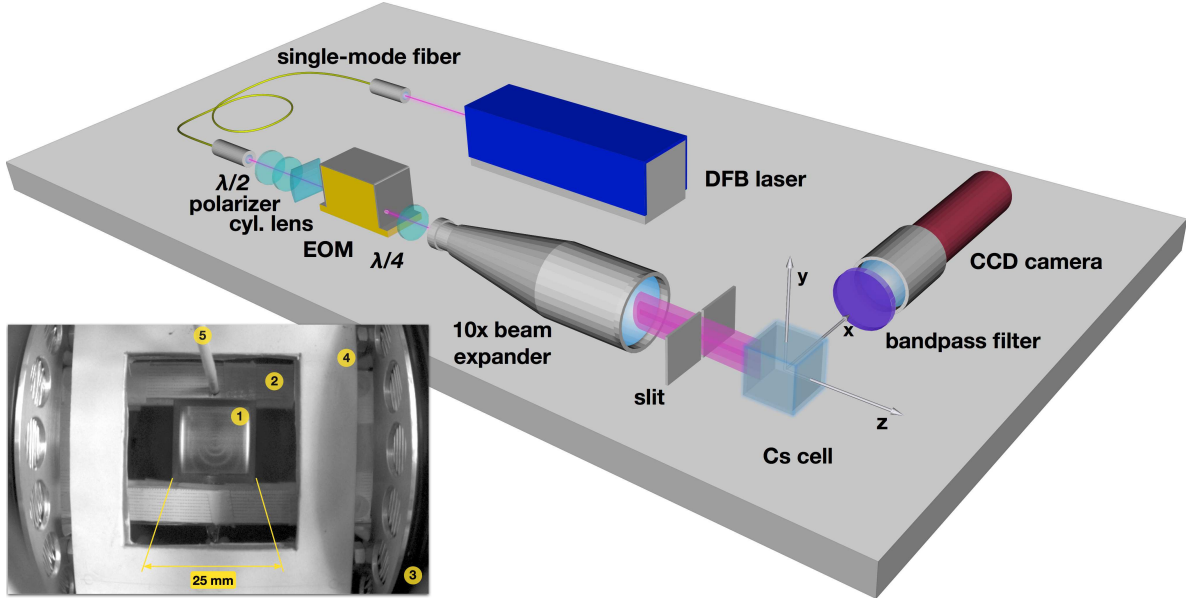}}
       \caption{\label{fig:Setup}  Schematic layout of the apparatus, not including magnetic field coils.
       The photograph shows the central part of the set-up, as seen by the CCD
       camera.
       One distinguishes: the cubic vapor cell (1) with ring-structured fluorescence produced by the 1WH field
       configuration, one of the support structures for the \textit{rf} coils (2), the coils producing the quadrupole field (3),
       a support structure for the 8 wires of the 8W-configuration (4), and the protruding `wire' of the 1WH field configuration.}
\end{figure*}

%%%%%%%%%%%%%%%%%%%%%%%%%%%%%%%%%%%%%%%%%%%%%%%%%%%%%%%%%%%%%%%%%%%%%%%%%%%%%%%%%%%%%%
%
Figure \ref{fig:Setup} shows a schematic of the experimental apparatus.
In what follows we will refer to a coordinate system, in which the laser
beam direction $\vec{k}$ is along $\hat{z}$, the $x$-axis is
along the horizontal direction perpendicular to $\vec{k}$, and
$\hat{y}$ is along the vertical direction, the origin of coordinates being chosen in the center of the spectroscopy cell.

The experiments used 894~nm radiation from a distributed feedback (DFB) laser, whose beam was carried by a single-mode fiber into a optical set-up.
The intensity of the collimated  beam at the fiber output
was $\approx$ 1~mW.
In principle, the frequency stability of the DFB laser was sufficient to carry out the measurements without additional stabilization, but for better performance we actively stabilized the laser frequency to the $F{=}4\rightarrow F'{=}3$ hyperfine transition of the cesium $D_1$ line.

 The collimated output beam from the fiber was prepared linear by a combination of $\lambda/2$-plate, polarizer and $\lambda/4$-plate allowing independent control of the laser power and its polarization.
For the Bell-Bloom experiment requiring $\sigma_+/\sigma_-$-modulation we inserted an electro-optic modulator
(Thorlabs, model EO-AM-NR-C1 with model HVA200 driver) before the $\lambda/4$-plate.
The EOM was driven  by a square wave
delivered by a  (Keithley, model 3390) waveform generator.
The laser beam profile was stretched to an elliptical shape by a cylindrical lens
placed after the polarizer, and was then expanded 10$\times$ by a spherical lens telescope.
Before entering the spectroscopy cell, a narrow
vertical strip of the expanded profile was defined by a 1.2 mm wide slit.
The beam entering the cell thus had sharply defined boundaries in
the direction of observation, while its vertical ($y$) intensity
profile was a Gaussian with a FWHM of 15~mm.
The spectroscopy cell was a cubic Pyrex cell (inner volume of
22${\times}$22${\times}$22~mm$^3$) with five optical quality
windows containing saturated Cs vapor at ambient temperature
together with a mixture of Ar (8~mbar) and Ne (45~mbar ) buffer gas,
which confined the Cs atoms.

In an auxiliary experiment we have determined the spin coherence
relaxation rate in a transmission ground state Hanle experiment
(similar to the one described by in Ref.~\cite{Castagna2011}),
yielding a 2~kHz (HWHM) linewidth.
As discussed in Sec.~\ref{sec:summary}, that linewidth becomes
significantly narrower when the experiments are performed in a
magnetically shielded environment.
%
%The
%fluorescence from our buffer gas cell came completely depolarized in
%used range of value of magnetic fields.

\looseness-1Three mutually orthogonal pairs of (300${\times}$300~mm$^2$) square Helmholtz coils were used to zero all components of the earth magnetic field.
The \textit{rf} field was produced by two rectangular (44${\times}$100~mm$^2$) coils spaced by 22~mm driven by single sine wave or a superposition of sine waves delivered by a programmable (Keithley, model 3390) function generator.
When using multi-component \textit{rf} fields, we accounted for the frequency-dependent impedance of the \textit{rf}-coil by giving the individual frequency components appropriate voltage amplitudes such as to deliver identical currents to the coils.

%%%%%%%%%%%%%%%%%%%%%%%%%%%%%%%%%%%%%%%%%%%%%%%%%%%%%%%%%%%%%%%%%%%%%%%%%%%%%%%%%%%%%%
%\subsection{\label{sec:FieldConfig}Magnetic field configurations}
%%%%%%%%%%%%%%%%%%%%%%%%%%%%%%%%%%%%%%%%%%%%%%%%%%%%%%%%%%%%%%%%%%%%%%%%%%%%%%%%%%%%%%
%

%
For the magnetic field mapping experiments reported below we have produced three types of inhomogeneous fields using three distinct coil and wire configurations as shown in Fig.~\ref{fig:CoilConfig}.

\subsection{Quadrupole ($Q$) field configuration}
\begin{figure*}[htbp]
    \centering
       \resizebox{\textwidth}{!}{\includegraphics{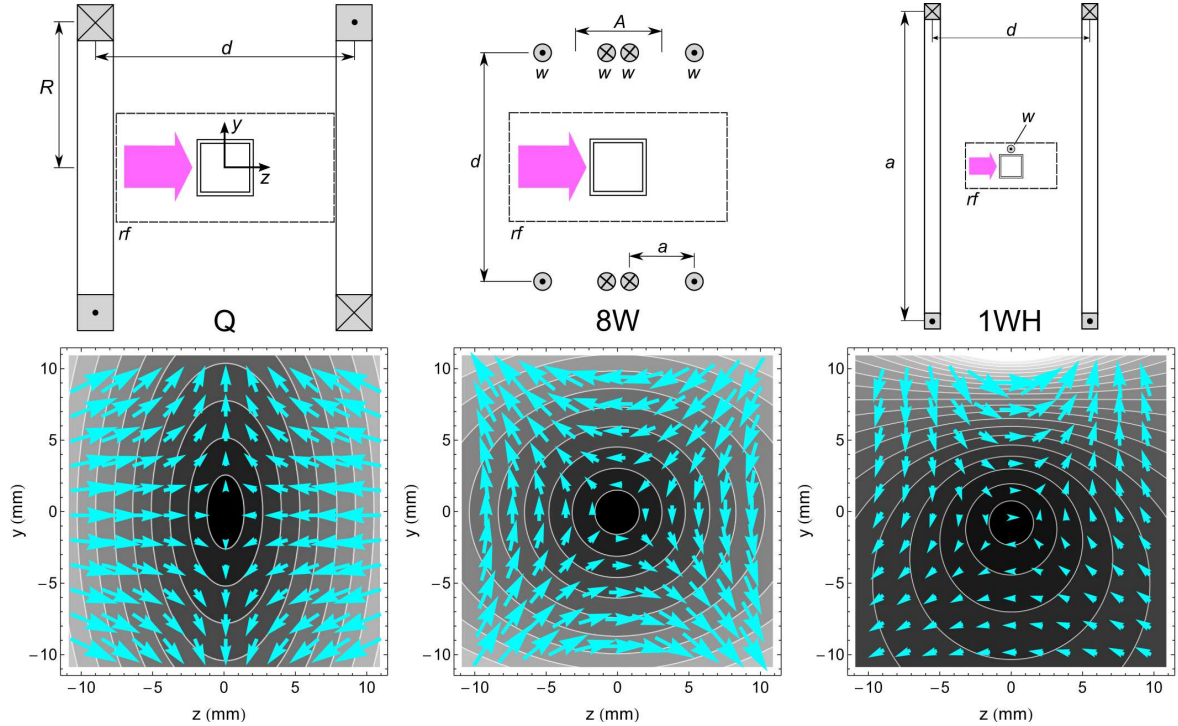}}
       \caption{\label{fig:CoilConfig}
       Top row: The three magnetic field configurations studied in this work, shown as cross sections in the $x{=}0$
       plane.
       From left to right: Quadrupole (Q), eight wire (8W), and single wire plus Helmholtz (1WH) field
       configurations.
       The relative dimensions for each configuration are to scale,
       except for the wires ($w$) whose diameters are magnified 3 times.
       The cubic spectroscopy cell is located in the center of each coil arrangement
     The dashed rectangles represent the two \textit{rf}-coils that are offset from the $x{=}0$ plane by $\pm$~11~mm.
     The bottom row shows the calculated contour plots of the magnetic scalar potential in the spatial region covered by the cell, together with superimposed vector plots of the magnetic field lines in the $x=0$ plane.}
\end{figure*}

Two circular coils, are excited by identical currents of opposite
sign, thus producing  a quadrupole field that is well described by
$\vec{B}(\vec{r})=B_0\,(x\,\hat{e}_x+y\,\hat{e}_y-2z\,\hat{e}_z)$,
with $\vec{r}=(x,y,z)$ over the cell dimension.
The top graph of the left column in Fig.~\ref{fig:CoilConfig} shows the $Q$-configuration, together with the spectroscopy cell.
The bottom of the same column shows the magnetic field lines
represented by arrows in the $x=0$ plane (the plane imaged by the
CCD camera looking along the $x$ direction), the arrow heads' size
representing the magnetic field strength.
The field lines are superposed on a contour plot of $|\vec{B}|$; the
scalar magnetic equipotential surfaces being ellipsoids with aspect
ratios $2{:}1$ in the $y{-}z$ and $x{-}z$ coordinates;
the equipotential lines in the $x{=}0$ plane are thus ellipses with the same aspect ratio
The characteristic feature of the $Q$-configuration are constant field gradients $dB_z/dz$ and $dB_y/dy$ with
magnitudes in the ratio $-2{:}1$ along the $z$- and $y$-axes,
respectively.

\subsection{Eight-wire (8W) field configuration}
Eight long wires arranged as shown in the top of the
central column in Fig.~\ref{fig:CoilConfig} produce homogeneous
gradients $dB_y/dz$ and $dB_z/dy$ of equal magnitude, as visualized
by the field lines in the lower part of the figure.
In the experimental set-up, the wires  were 90~mm long copper rods.
The iso-potential lines of the 8W-configuration in the
$x{=}0$ plane in the vicinity of $y{=}0$ and $z{=}0$ are circles.

\subsection{Helmholtz plus single wire  (1WH) field configuration}
The third field configuration consisted of two square
coils in Helmholtz configuration, producing a homogeneous field
along $\hat{z}$, onto which the field produced by a single current-carrying wire
($w$) along $\hat{x}$ is superposed.
The sign of the wire current was chosen such as to compensate the
homogeneous field at a point near the center of the cell.
Such a field configuration, known as `z-wire trap' is commonly used
for the magnetic trapping of cold atoms on atom chips (see,
e.g.,~\cite{Folman2002}) since changing the relative magnitude of
the wire and Helmholtz coil currents allows the controlled
displacement of the zero field point along the vertical ($y$)
direction.
The coil configuration and a typical field pattern of its non-concentric
potential lines are shown in the right column of
Fig.~\ref{fig:CoilConfig}.
%

%

%%%%%%%%%%%%%%%%%%%%%%%%%%%%%%%%%%%%%%%%%%%%%%%%%%%%%%%%%%%%%%%%%%%%%%%%%%%%%%%%%%%%%%
\subsection{Data recording and analysis}
%%%%%%%%%%%%%%%%%%%%%%%%%%%%%%%%%%%%%%%%%%%%%%%%%%%%%%%%%%%%%%%%%%%%%%%%%%%%%%%%%%%%%%
The laser-induced fluorescence from the irradiated Cs vapor layer
was imaged by a 16-bit CCD camera (model ST-i monochrome from SBIG
Astronomical Instruments, 640$\times$480 pixels), whose
line-of-sight was directed along $\hat{x}$ (Figure \ref{fig:Setup}).
The zoom lens (ARBUS, model TV8553) was adjusted such that an area
slightly larger than the (outer) 25${\times}$25~mm$^2$ cross section
of the cell was imaged onto the CCD sensor.
The LIF intensity variation of interest covered $\approx$10\% of the 65'536 grey
tones provided by the camera, because of bright regions of scattered light from the lateral windows, as can be seen in the photograph insert of Fig.~\ref{fig:Setup}.
We note that with a conventional 8-bit camera that variation would
cover only $\approx$25 shades of grey.
An 894~nm interference filter mounted on the imaging lens suppressed
stray room light.

The typical data recording procedures were as follows:
In experiments involving depolarization by \textit{rf}-irradiation we recorded two images with the same exposure
time of typically 20--30~s, viz., one image (main image) with the magnetic field
gradient and the \textit{rf}-field applied to the cell, and one image (reference image)
without applied \textit{rf}-field.
For recording the reference image the gradient field was switched off and replaced by a polarization stabilizing field $\beta_\parallel{\gg}1$, oriented along $\hat{k}$ for experiments on oriented atoms, and along the light polarization for experiments on aligned atoms.
In an off-line analysis both images were cropped to display only the inner cell cross section of 22${\times}$22~mm$^2$, after which the reference image was subtracted from the main image.
The grey tones in the difference image were then rescaled to span the full range from white to black by assigning 0 to the darkest pixel and 1 to the brightest pixel and mapping all intermediate grey tones in a linear manner onto the [0,1] interval.

Based on Eq.~\eref{eq:Pf10Final}
the differential fluorescence image of \textit{oriented} atoms is thus described by the expression
\begin{eqnarray}\label{eq:deltaPf10}
\delta P_{f}^{(1)}(\vec{r})&=&P_{f}^{(1)}(\vec{r}, S_{rf},\beta_\parallel,\beta_\perp)-P_{f}^{(1)}(\vec{r}, S_{rf}{=}0,\beta_\parallel{\gg}1,\beta_\perp{=}0)\nonumber\\
&\propto &C^{(1)}\left[\frac{1+\beta_\parallel^2(\vec{r})}{1+\beta_\parallel^2(\vec{r})+\beta_\perp^2(\vec{r})}\,
\left(1-\frac{\sqrt{S_{rf}}}{[\beta_{rf}-\beta(\vec{r})]^2+S_{rf}+1}\right)\right],
\end{eqnarray}
where the local magnetic field components $\beta_\parallel$ and $\beta_\perp$, with  $\beta{=}\sqrt{\beta_\parallel^2+\beta_\perp^2}$ of each voxel at position $\vec{r}$ in the imaged vapor layer determine the fluorescence intensity.
Setting $C^{(1)}{=}1$ yields images with pixel intensities in the range $[0,1]$, the same range as covered by the experimental pictures.

%\pagebreak

\looseness-1In the same way, one derives for the differential fluorescence image of \textit{aligned} atoms
\begin{eqnarray}\label{eq:deltaPf20}
\fl \delta P_{f}^{(2)}(\vec{r})&=&P_{f}^{(2)}(\vec{r}, S_{rf},\beta_\parallel,\beta_\perp)-P_{f}^{(2)}(\vec{r}, S_{rf}{=}0,\beta_\parallel{\gg}1,\beta_\perp{=}0)\nonumber\\
\fl &\propto &C^{(2)}\left[1{-}\left(\frac{1}{4}{+}
\frac{3}{4}\frac{1{+}8\beta_\parallel^2{+}16\beta_\parallel^4}{1{+}4\beta_\parallel^2{+}4\beta_\perp^2}
{-}3\frac{\beta_\parallel^2{+}\beta_\parallel^4}{1{+}\beta_\parallel^2{+}\beta_\perp^2}\right)%\times\nonumber\\
\left(1{-}\frac{\sqrt{S_{rf}}}{(\beta_{rf}{-}\beta)^2{+}S_{rf}{+}1}\right)\right]\,,
\end{eqnarray}
where, as above, $C^{(2)}{=}1$ yields images with pixels values spanning the range from 0 (black) to 1 (white).

In the experiment described in Sec.~\ref{sec:CircPolBonly} that did
not use an \textit{rf} magnetic field, the reference image was
recorded by replacing the inhomogeneous magnetic field with
a polarization-stabilizing longitudinal field
$\beta_\parallel{\gg}1{\gg}\beta_\perp$.
Using Eq.~\eref{eq:Pf10All}, one easily sees that the difference image in that case is given by
\begin{eqnarray}\label{eq:deltaPf10InhomOnly}
\delta P_{f}^{(1)}(\vec{r})&=&P_{f}^{(1)}(\vec{r}, \beta_\parallel,
\beta_\perp)-P_{f}^{(1)}(\vec{r}, \beta_\parallel{\gg}1,
\beta_\perp{=}0)\nonumber\\
&\propto &\frac{\beta_\perp^2(\vec{r})}{1+\beta_\parallel^2(\vec{r})+\beta_\perp^2(\vec{r})}\,.
\end{eqnarray}

The experimental results in the following sections have been modeled
by Eqs.~(\ref{eq:deltaPf10}--\ref{eq:deltaPf10InhomOnly}), by adjusting only the \textit{rf} saturation parameter $S_{rf}$ and the spin relaxation rate $\gamma$.
In order to account for the inhomogeneous vertical intensity
distribution of the laser beam in the cell, the theoretical
fluorescence patterns in
Eqs.~(\ref{eq:deltaPf10}--\ref{eq:deltaPf10InhomOnly}) were
multiplied by a Gaussian exp$[-y^2/2\sigma_y^2]$ with a FWHM of
15~mm.

%%%%%%%%%%%%%%%%%%%%%%%%%%%%%%%%%%%%%%%%%%%%%%%%%%%%%%%%%%%%%%%%%%%%%%%%%%%%%%%%%%%%%%
\section{\label{sec:ExpResults}Experimental results}
%%%%%%%%%%%%%%%%%%%%%%%%%%%%%%%%%%%%%%%%%%%%%%%%%%%%%%%%%%%%%%%%%%%%%%%%%%%%%%%%%%%%%%
%
%%%%%%%%%%%%%%%%%%%%%%%%%%%%%%%%%%%%%%%%%%%%%%%%%%%%%%%%%%%%%%%%%%%%%%%%%%%%%%%%%%%%%%
\subsection{\label{sec:CircPolBonly}Fluorescence of oriented atoms in inhomogeneous magnetic fields}
\begin{figure*}[tb]
    \centering
       \resizebox{\textwidth}{!}{\includegraphics{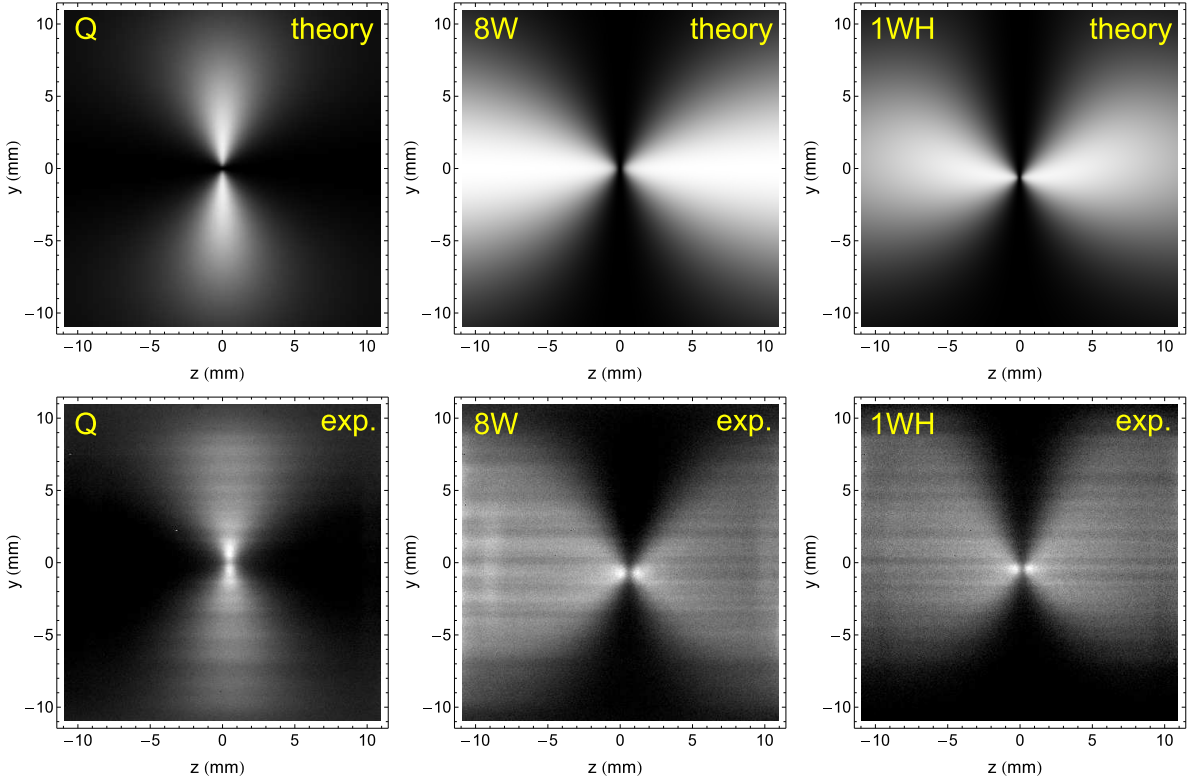}}
       \caption{\label{fig:CircPolNoRF} Two-dimensional fluorescence distributions induced by circularly polarized laser light in the three inhomogeneous field configurations shown in Fig.~\ref{fig:CoilConfig}.
       From left to right: quadrupole field ($Q$),  8-wire field ($8W$), and Helmholtz plus one-wire field ($1WH$).
       The top row represents the LIF distribution calculated with Eq.~\eref{eq:Pf10All} and the lower row the experimentally recorded fluorescence patterns.
       }
\end{figure*}

%%%%%%%%%%%%%%%%%%%%%%%%%%%%%%%%%%%%%%%%%%%%%%%%%%%%%%%%%%%%%%%%%%%%%%%%%%%%%%%%%%%%%%
%
The bottom row of Fig.~\ref{fig:CircPolNoRF} shows fluorescence images from the cell excited by circularly polarized light in the three inhomogeneous fields configurations described in Sec.~\ref{sec:setup}.
We refer to these fluorescence patterns as `Hanle background fluorescence' images.
The dark horizontal stripes on the fluorescence images are likely due to an interference effect from scattered laser light. Alternatively, the stripes could be shadows or reflections that are produced by small metallic Cs droplets on the windows of the cell.

\begin{figure*}[htbp]
    \centering
       \resizebox{\textwidth}{!}{\includegraphics{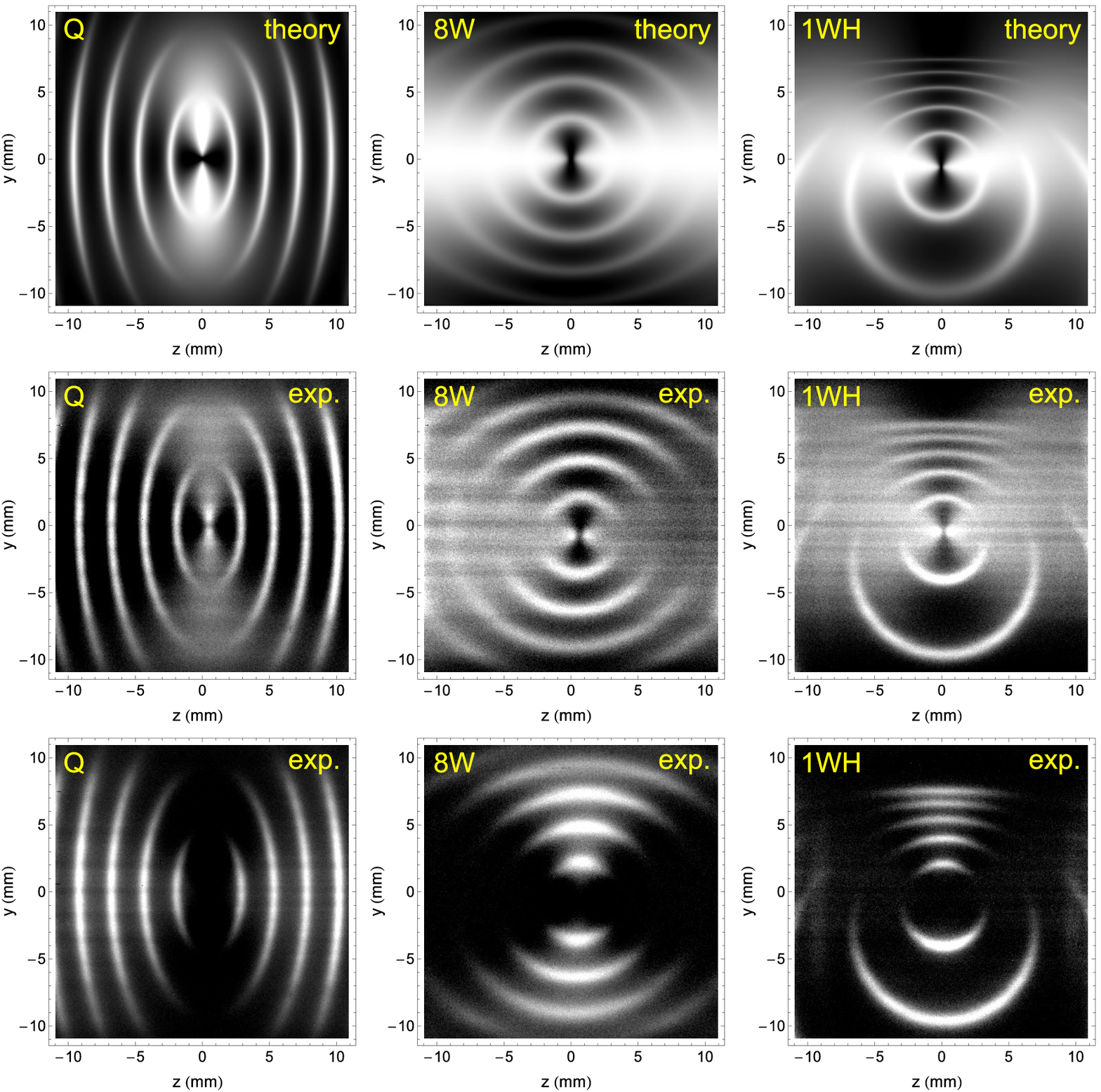}}
       \caption{\label{fig:CircPolWithRF} Magnetic resonance depolarization of fluorescence induced by \textit{circularly-polarized} light in the three inhomogeneous field configurations.
       A multi-component \textit{rf}-field is used space-selectively
       depolarizes the atoms, and the bright lines represent scalar magnetic
       iso-potential lines.
        For the $Q$- and the $1WH$- configurations the bright ellipses and circles represent--from inside to outside--lines along which the magnetic field modulus $\beta$ varies
        from 19
        %19, 37, 56, 73 and 93 $\mu$T,
        to 93~$\mu$T in steps of 18.5~$\mu$T.
For the $8W$-configuration the corresponding fields were
%10, 20, 30, 40 and 50
10\dots50 $\mu$T (10~$\mu$T steps).
       The top row represents the LIF distributions predicted by Eq.~\eref{eq:deltaPf10} and the middle row the corresponding measured differential fluorescence patterns.
       The graphs in the bottom row were produced by subtracting a reference image recorded by switching off the \textit{rf} field, while leaving the inhomogeneous field applied.}
\end{figure*}

%The recorded pixel intensities are scaled such that the brightest
%and darkest pixels correspond to $1$ (white) and $0$ (black),
%respectively.
%
In the top row we show the corresponding modeled images predicted by
Eq.~\eref{eq:deltaPf10InhomOnly} using the spatial field
distributions $\beta_\parallel(\vec{r})$ and $\beta_\perp(\vec{r})$
of the three coil configurations, noting that the subscripts
$\parallel$ and $\perp$ refer to the local magnetic field being
parallel or perpendicular to the $\hat{k}$-vector of the exciting
light beam.
The observed and modeled fluorescence patterns show an excellent
agreement, except near $\vec{r}=0$, where the dark spot is slightly
less extended in the experimental images, a fact that we assign to
laboratory field gradients, noting that, after all, the experiments
are carried out in a magnetically unshielded environment.
A small contribution from atomic alignment, not taken into account
in our model, may also contribute to the slight deviation near
$\vec{r}{=}0$.

%%%%%%%%%%%%%%%%%%%%%%%%%%%%%%%%%%%%%%%%%%%%%%%%%%%%%%%%%%%%%%%%%%%%%%%%%%%%%%%%%%%%%%
\subsection{\label{sec:CircPolBplusRF}Fluorescence of rf-depolarized oriented atoms in inhomogeneous magnetic field}
%%%%%%%%%%%%%%%%%%%%%%%%%%%%%%%%%%%%%%%%%%%%%%%%%%%%%%%%%%%%%%%%%%%%%%%%%%%%%%%%%%%%%%
%
%
Although the results presented in the previous paragraph show a good
agreement between observations and the forward model predictions,
they are not very useful for addressing the inverse problem that
consists in inferring unknown  magnetic field distributions from the
fluorescence patterns that they produce.

As discussed in Sec.~\ref{sec:rfDepol}, the injection of a weak
oscillating magnetic field (\textit{rf}-field) into the cell is a
powerful tool for addressing this inverse problem.
Magnetic resonance transitions induced by the \textit{rf}-field
oscillating at $\omega_{rf}$ will depolarize atoms at
locations where the magnetic resonance condition
$\beta{=}|\vec{\beta}|{=}\sqrt{\beta_\parallel^2+\beta_\perp^2}{\equiv}\omega_{rf}/\gamma_F$
is fulfilled.
This depolarization leads to an enhanced fluorescence
intensity, thus highlighting regions of constant magnetic field
strength ($\beta${=}const), i.e., yielding bright scalar
iso-potential lines.
In order to demonstrate this approach we have exposed the cell to an
\textit{rf}-field consisting of a comb of 5 equidistant harmonic
oscillations.
%
%Based on a measurement of the complex impedance $Z(\omega)$ of the
%rf coils, the relative amplitudes (voltage) of the frequency
%components were adjusted to yield identical coil currents at each
%frequency.
%

\begin{figure}[htbp]
\begin{center}
\includegraphics[width=\columnwidth]{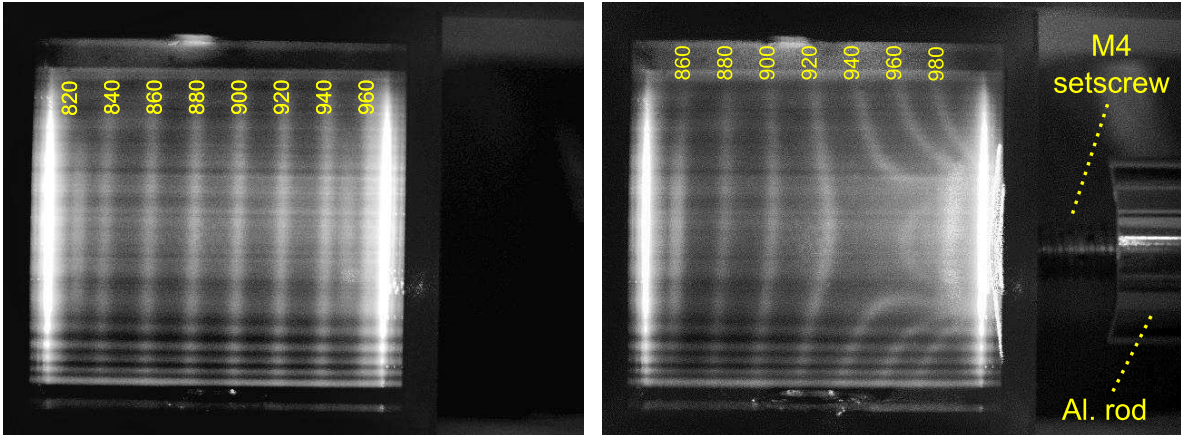}
\end{center}
 \caption{\label{fig:setscrew}
        Potential lines in a region with a homogeneous field
       gradient (left) and deformation of the potential by an M4-setscrew (right).}
\end{figure}

For the $Q$- and the $1WH$- configurations the \textit{rf}-field
consisted of a fundamental oscillation at
$\omega_{rf}/2\pi{=}$65~kHz and 4 harmonics thereof, while
the measurements with the $8W$-field involved the fundamental and
harmonics of 35~kHz.
The graphs in the middle row of Fig.~\ref{fig:CircPolWithRF} show
the experimental results, in which the bright iso-potential lines
clearly stand out against the smooth Hanle background fluorescence
pattern.
The fluorescence predicted by Eq.~\eref{eq:deltaPf10} is shown in the top row of Fig.~\ref{fig:CircPolWithRF}.
The results in the top two rows were chosen to illustrate that our model calculations do not only reproduce the bright potential lines, but also the underlying Hanle background fluorescence patterns.

For practical applications one may wish to suppress the Hanle background.
This can be achieved by recording the reference image with merely switching off the \textit{rf} field, and leaving the inhomogeneous field applied.
The corresponding results are shown in the bottom row of  Fig.~\ref{fig:CircPolWithRF}.

As an application we show in Fig.~\ref{fig:setscrew} the
deformations of the potential lines by a magnetized
stainless steel setscrew.
The left graph shows a set of iso-field lines in a field produced by
the superposition of the (homogeneous) field from a pair of
Helmholtz coils and a small linear $dB_z/dz$-gradient field
produced by the quadrupole coil $Q$.
The \textit{rf}-field for that measurement consisted of a
superposition of oscillations at frequencies ranging from 820 to
960~kHz, yielding bright fluorescence lines at fields $B_z$ ranging
from 234.3 to 274.3~$\mu$T, in steps of 5.7~$\mu$T.

When an M4 stainless steel setscrew held by a non-magnetic aluminum
rod is approached to the cell from the right side, the iso-potential
lines are displaced and deformed as shown on the right graph of
Fig.~\ref{fig:setscrew}.
We show this example merely as a qualitative demonstration of the
method.
Work on quantitative methods for inferring the field produced by the
perturbing object is in progress.

\begin{figure*}[htbp]
    \centering
       \resizebox{\textwidth}{!}{\includegraphics{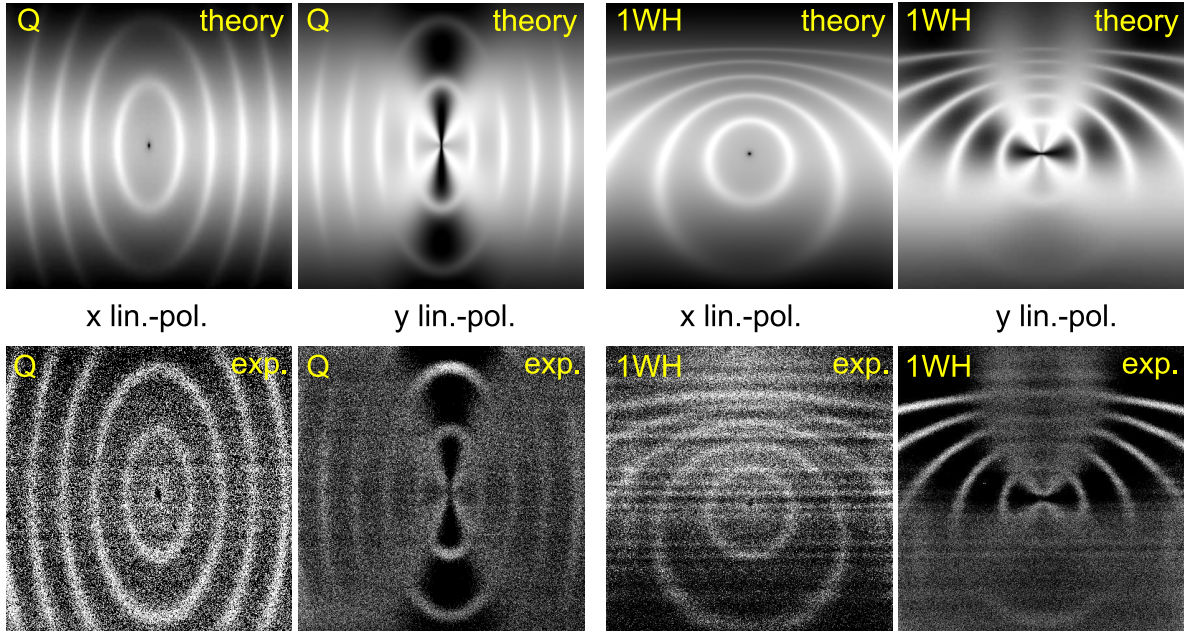}}
       \caption{\label{fig:LinPolWithRF}
       Magnetic resonance depolarization of fluorescence induced by \textit{linearly-polarized} light in fields produced by the quadrupole ($Q$) and one wire plus Helmholtz ($1WH$) coils, respectively.
       Fluorescence images are shown for laser polarization parallel ($x$) and perpendicular ($y$) to the camera's line of sight.
       A multi-component \textit{rf}-field consisting of the fundamental and harmonics of 65~kHz  is used to space-selectively depolarize atoms.
       The top row represents the LIF distributions calculated with Eq.~\eref{eq:deltaPf20} and the lower row the recorded differential fluorescence pattern.}
\end{figure*}

%%%%%%%%%%%%%%%%%%%%%%%%%%%%%%%%%%%%%%%%%%%%%%%%%%%%%%%%%%%%%%%%%%%%%%%%%%%%%%%%%%%%%%
\subsection{\label{sec:LinPolBplusRF}Fluorescence from rf-depolarized aligned atoms in inhomogeneous magnetic field}
%%%%%%%%%%%%%%%%%%%%%%%%%%%%%%%%%%%%%%%%%%%%%%%%%%%%%%%%%%%%%%%%%%%%%%%%%%%%%%%%%%%%%%
%
We have also recorded fluorescence images induced by
\textit{linearly-polarized} light that creates an atomic alignment
that is then resonantly destroyed by the rf-field in a
space-selective manner.
The use of linear polarization introduces an additional degree of
freedom, viz., the relative orientation of the direction of
polarization and the axis of symmetry  (if any) of the gradient
producing coils.

Results and their corresponding simulation based on
Eq.~\eref{eq:deltaPf20} are shown in Fig.~\ref{fig:LinPolWithRF} for
the quadrupole ($Q$) field  and the one-wire plus Helmholtz ($1WH$)
field.
We find again an excellent agreement between the non-trivial simulated and measured patterns.
We note in particular the dark spot in the zero-field point that is particularly well visible in the experiments with $x$-polarized light.
We also note the poorer signal/noise ratio of the experimental data
compared to the data obtained with circularly polarized light.
This is due to the fact that the buffer gas reduces the optical
pumping efficiency with linearly polarized light.

The dependence of the fluorescence maps on the orientation of the light polarization can be understood in a quantitave manner as follows.
For the $Q$-configuration excited, e.g., with $y$-polarized light
Fig.~\ref{fig:CoilConfig} shows that there is a growing
alignment-stabilizing field $B_y$ along the $y$-direction, hence the
dark band of growing width that develops as one moves away (along
$y$) from the center of the cell.
If, on the other hand, the light polarization (in the same $Q$-configuration) is along $x$, all field components in the imaged $x=0$ plane (except for the spot at $y{=}z{=}0$) have a depolarizing effect, hence the overall brighter image.
Similar arguments allow one to understand the dark and bright zones
in the $1WH$-configuration.

%%%%%%%%%%%%%%%%%%%%%%%%%%%%%%%%%%%%%%%%%%%%%%%%%%%%%%%%%%%%%%%%%%%%%%%%%%%%%%%%%%%%%%%%%%%%%%%%%%%%%%%%%%%%%%
\subsection{\label{sec:DarkLineImaging}Dark resonance potential imaging by helicity-modulated light}
%%%%%%%%%%%%%%%%%%%%%%%%%%%%%%%%%%%%%%%%%%%%%%%%%%%%%%%%%%%%%%%%%%%%%%%%%%%%%%%%%%%%%%%%%%%%%%%%%%%%%%%%%%%%%%
%
In 1961 Bell and Bloom have shown~\cite{Bell61}  that a beam of
intensity-modulated light induces magnetic resonance transitions in
alkali atoms exposed to a transverse magnetic field when the
modulation frequency matches the Larmor precession frequency.
Recently we have extended the Bell-Bloom technique to polarization
modulation \cite{Grujic2013,Fescenko2013} in a transverse magnetic
field.
When the helicity $\xi$ of a circularly-polarized light beam is
periodically reversed in synchronicity with the precession of the
medium's spin polarization, a resonant build-up of the polarization
occurs when the modulation frequency $\omega_{mod}$ matches an
integer multiple of the Larmor frequency $\omega_L$.
This build-up manifest itself as a resonant reduction of the
medium's absorption coefficient when $\omega_{mod}=n\omega_L$, and
hence a corresponding increase of the transmitted light power~\cite{Fescenko2013} .
\begin{figure*}[tpb]{!}
    \centering
       \resizebox{\textwidth}{!}{\includegraphics{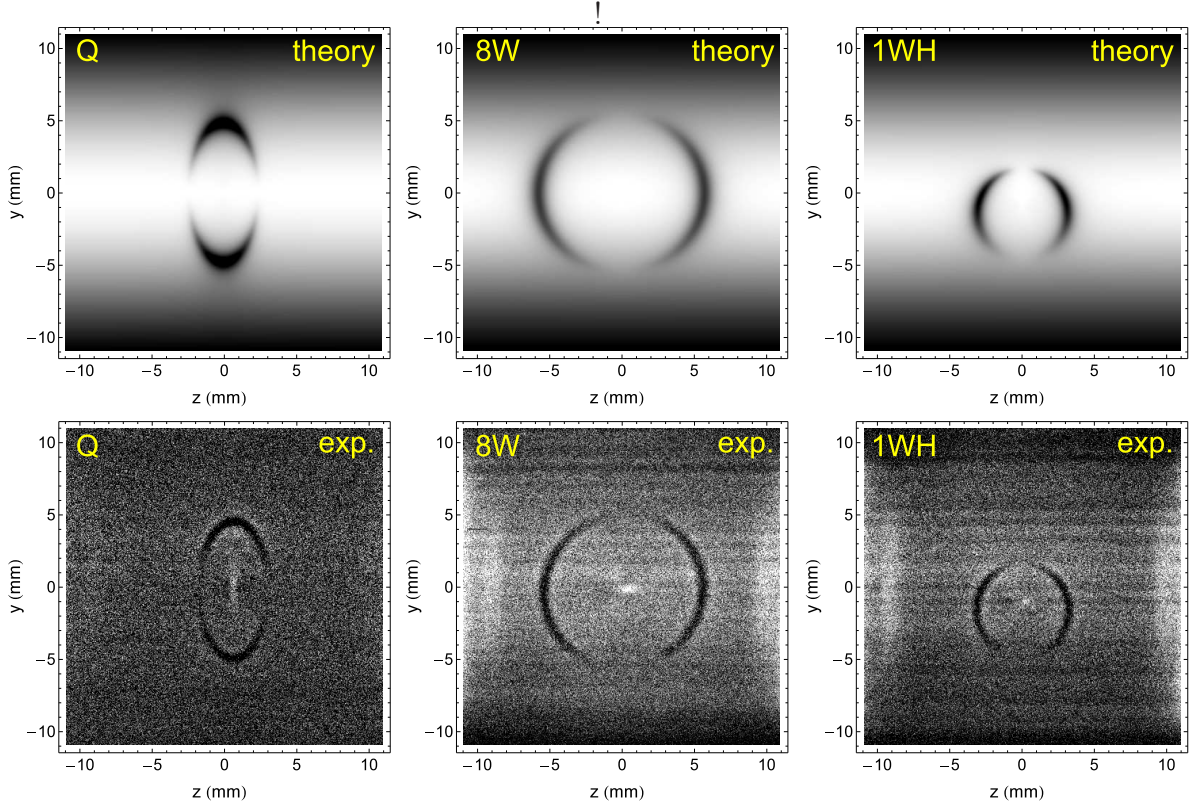}}
       \caption{\label{fig:BellBloom}
       Fluorescence images showing a dark magnetic potential line (superposed on the Hanle background fluorescence) induced by the resonant creation of spin polarization
       using a light beam with \emph{circular polarization modulation}.
       Experiments (bottom) were carried out in the three inhomogeneous field configurations $\nu_{mod}{=}$65~kHz in the $Q$- and
       $1WH$-fields, and 35~kHz in the $8W$-field.
       The top row shows the differential LIF distribution calculated by Eq.~\eref{eq:deltaPf10BellBloom}.}
\end{figure*}

A reduction of the medium's fluorescence is expected to go in pair
with this variant of the EIT (electromagnetically-induced
transparency) phenomenon.
The graphs in the bottom row of Fig.~\ref{fig:BellBloom} show fluorescence images recorded with circular polarization-modulated light in our three standard field configurations.
In these recordings the reference image was recorded by switching off the polarization modulator, leaving the light in an unknown (i.e., unmeasured) elliptical polarization state.
Since we did not (as for the recordings of Figs.\ref{fig:CircPolWithRF}, \ref{fig:LinPolWithRF}) turn off the gradient field for recording the reference images, the Hanle fluorescence background is suppressed in the differential images.

The experimental images show well the anticipated dark potential line, at positions where the Larmor frequency matches the polarization modulation frequency.
With $\omega_{mod}/2\pi{=}$65~kHz the dark line occurs at a field of
19~$\mu$T in the $Q$- and $1WH$-configurations, while it occurs at
10~$\mu$T in the $8W$-configuration for which $\omega_{mod}/2\pi$
was 35~kHz.

On notes the much poorer signal contrast of the (all-optical)
modulation experiment compared to the experiments involving
\textit{rf}-depolarization that is partially due to the square-wave
polarization modulation, of which only the fundamental Fourier
component contributes.
The graphs in the top row of of Fig.~\ref{fig:BellBloom} show images that were modeled by the function
\begin{eqnarray}\label{eq:deltaPf10BellBloom}
\delta P_{f}^{(1)}(\vec{r})
\propto 1-\frac{\beta_\parallel^2(\vec{r})}{1+\beta_\parallel^2(\vec{r})+\beta_\perp^2(\vec{r})}
\times\frac{1}{[\beta_{mod}-\beta(\vec{r})]^2+1} \,.
\end{eqnarray}
applied to all three gradient configurations.
In the last expression we have retained
only the fundamental Fourier component of the polarization
modulation function, since the light intensity was too low to excite the harmonic resonances that are expected to occur at positions $\vec{r}$, where $\omega_{mod}=(n{>}1)\omega_L(\vec{r})$.
These Bell-Bloom dark resonances are complementary to the bright
resonances observed with \textit{rf}-depolarization
(Fig.~\ref{fig:CircPolWithRF}).
%

%%%%%%%%%%%%%%%%%%%%%%%%%%%%%%%%%%%%%%%%%%%%%%%%%%%%%%%%%%%%%%%%%%%%%%%%%%%%%%%%%%%%%%%%%%%%%%%%%%%%%%%%%%%%%%
\section{\label{sec:SpatialResolution}Spatial resolution and magnetometric sensitivity of field mapping by fluorescence imaging}
%%%%%%%%%%%%%%%%%%%%%%%%%%%%%%%%%%%%%%%%%%%%%%%%%%%%%%%%%%%%%%%%%%%%%%%%%%%%%%%%%%%%%%%%%%%%%%%%%%%%%%%%%%%%%%
%%
We estimate the spatial resolution $\Delta z$ of the iso-potential lines as follows:
Suppose that spin-oriented atoms are in a local polarization stabilizing field, onto which a weak gradient field $G_{zz}{=}dB_z/dz$ is superposed.
The interaction with the \textit{rf}-field will then depolarize atoms, i.e., make the fluorescence light up in a region given by $\Delta z=\delta B/G_{zz}$, where $\delta B$ is the linewidth of the magnetic resonance line.
Increasing the gradient $G_{zz}$ will narrow the iso-potential lines, until the region containing depolarized atoms broadens due to the depolarized atoms diffusing out of the depolarization region proper.
In very strong gradients the spatial resolution will this be limited by the distance $\sqrt{2\,D\,\tau}$ over which depolarized atoms diffuse during their spin relaxation time~$\tau$, $D$ being the diffusion constant. 
Since the two effects are statistically independent, we add them quadratically to obtain
\begin{equation}\label{eq:SpatialResolution0}
\Delta z=\sqrt{2\,D\,\tau+\left(\delta B/G_{zz}\right)^2}\,.
 %
%  =\sqrt{2\,D\,T_1+\left(\frac{\delta\nu}{\widetilde{G}_{zz}}\right)^2}\,,
\end{equation}
We note that a similar approach was used by Tam~\cite{Tam1979}, who added the two contributions linearly.
Inspired by the latter work, we assume the spin relaxation rate to be given by 
\begin{equation}\label{eq:SpatialResolution1}
\Gamma=\tau^{-1}=\Gamma_\mathrm{buffer}+\Gamma_\mathrm{pump}+\gamma_F\,\delta B_\mathrm{50Hz}
\end{equation}
where $\Gamma_\mathrm{buffer}$ is the depolarization rate due to collisions with buffer gas atoms, and $\Gamma_\mathrm{pump}$ the optical pumping, i.e., photon scattering rate.
In our unshielded environment laboratory magnetic fields oscillating at the line frequency, artificially broaden the depolarization region, since they jiggle the region in which the \textit{rf}-depolarization occurs by $\approx$1~mm, as shown below.
We parametrize this effect in terms of $\delta B_\mathrm{50Hz}=\sqrt{\langle B^2(t)\rangle_\mathrm{rms}}$ and the corresponding broadening is one to two orders of magnitude larger than $\Gamma_\mathrm{buffer}$.
Since the same broadening mechanism determines the width of the magnetic resonance line (an additional contribution from spin exchange collisions to the latter~\cite{Tam1979} being completely negligible), we can estimate the field gradient dependence of the spatial resolution to be given by
\begin{equation}\label{eq:SpatialResolution2}
\Delta z\approx\sqrt{\frac{2\,D}{\gamma_F\,\delta B_\mathrm{50Hz}+\Gamma_\mathrm{pump}}+\left(\frac{\gamma_F\delta B_\mathrm{50Hz}+\Gamma_\mathrm{pump}}{G_{zz}}\right)^2}\,,
  %
%  =\sqrt{2\,D\,T_1+\left(\frac{\delta\nu}{\widetilde{G}_{zz}}\right)^2}\,,
\end{equation}
where it is reasonable to assume, although not explicitly proven, that optimal resolution conditions are achieved in the present configuration when $\Gamma_\mathrm{pump}$ is on the order of the 50~Hz broadening term.

\begin{figure}[htpb!]
    \centering
       \resizebox{0.75\columnwidth}{!}{\includegraphics{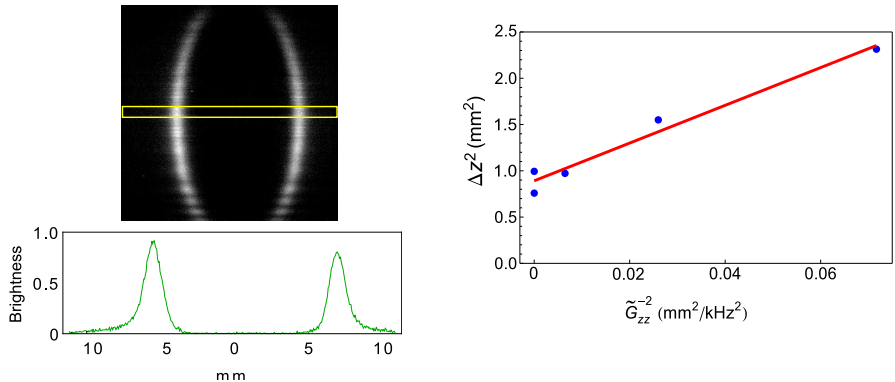}}
       \caption{\label{fig:SpatialRes}
      Dependence of the spatial resolution of the fluorescence field imaging method on the applied magnetic field gradient.
      Left: Typical fluorescence image with single \textit{rf}-depolarization in $Q$-gradient field
      and intensity profile from data in yellow box above.
      Right: Gradient dependence of the profile widths, together with linear fit.
      }
\end{figure}

Figure \ref{fig:SpatialRes} represents an experimental recording of
the spatial resolution's gradient dependence.
For this measurement we varied the value of the gradient in the
$Q$-configuration using a single harmonic oscillation whose frequency  was
adjusted to produce (for each gradient value) a single equipotential
line, such as shown on the top left of Fig.~\ref{fig:SpatialRes}  .
Cuts through the fluorescence maps (bottom left part of the figure) were then used to infer the
 FWHM widths $\Delta z$ of the bright potential lines.
The right graph of Fig.~\ref{fig:SpatialRes} shows the gradient dependence of $(\Delta z)^2$ together with a fit using the model function of 
Eq.~\eref{eq:SpatialResolution2}.
The intercept at strong gradients yields $\Delta z(G_{zz}{\rightarrow}\infty)$=0.94(4)~mm.
From the slope of the fitted line we infer an effective depolarization rate $\Gamma/2\pi$ of 4.5(2)~kHz, a
value that is compatible with the widths of magnetic resonance lines
measured in an auxiliary double resonance transmission experiment.

We estimate the magnetometric resolution of the method in its current implementations as follows:
The data in the cuts on the bottom left part of Fig.~\ref{fig:SpatialRes} represent fluorescence from a voxel of $\approx$ 1~mm$^3$, whose signal/noise ratio $SNR$ is $\approx$ 400.
We can thus state that the magnetometric sensitivity is $\delta
B_\mathrm{50Hz}/\gamma_F/SNR\approx$ 2~nT with a recording time of
20~s.

%%%%%%%%%%%%%%%%%%%%%%%%%%%%%%%%%%%%%%%%%%%%%%%%%%%%%%%%%%%%%%%%%%%%%%%%%%%%%%%%%%%%%%%%%%%%%%%%%%%%%%%%%%%%%%
\section{\label{sec:summary}Summary and outlook}
%%%%%%%%%%%%%%%%%%%%%%%%%%%%%%%%%%%%%%%%%%%%%%%%%%%%%%%%%%%%%%%%%%%%%%%%%%%%%%%%%%%%%%%%%%%%%%%%%%%%%%%%%%%%%
%
We have presented a tomographic method for mapping two-dimensional
distributions of magnetic scalar potentials that is based on the
position-selective destruction of spin-polarization by magnetic
resonance induced by a multi-component oscillating magnetic field.
Potential lines are directly visible as CCD camera images, and their contrast can be enhanced by subtraction of
suitable reference images.
We have also presented algebraic equations that allow the easy forward modeling of the fluorescence pattern for three distinct current geometries used to produce easily calculable inhomogeneous magnetic fields.

The proof-of-principle experiments were carried out in an unshielded environment, where they yield a spatial resolution on the order of 1~mm.
The resolution is limited by laboratory magnetic fields oscillating
at the line frequency which broaden the magnetic resonance lines to
a few kHz.
In view of porting the experiments into a magnetically shielded
environment we have already determined the magnetic resonance
linewidth of the same cell to be on the order of 50~Hz, when
recorded in a two-layer magnetic shield.
Based on this we expect the spatial resolution to be given by $\Delta z=\sqrt{D/\Gamma_\mathrm{pump}}$, neglecting spin-exchange relaxation and assuming optimal detection conditions ($\Gamma_\mathrm{pump}{\approx}\Gamma_\mathrm{buffer}$).
Based on this we believe that a spatial resolution on the order of 100 $\mu$m can be achieved when carrying out the experiments (using the same vapor cell) in a shielded environment.
Work towards such measurements is in progress.
In parallel we work on methods allowing the inverse reconstruction of spatially-resolved field patterns from sources with an unknown magnetization.

\ack{%%%%%%%%%%%%%%%%%%%%%%%%%%%%%%%%%%%%%%%%%%%%%%%%%%%%%%%%%%%%%%%%%%%%%%%%%%%%%%%%%%%%%%%%%%%%%%%%%%%%%%%%%%%%%
We acknowledge financial support by Scopes Grant IZ\,76Z0\_147548/1 from the Swiss National Science Foundation.
I.F. acknowledges support by Grant 11.133 of the Scientific Exchange Programme
(Sciex-NMS$^\mathrm{ch}$), and from FOTONIKA-LV FP7-REGPOT-CT-2011-285912 project.
We thank our colleagues Z.~Gruji\'c, V.~Lebedev, V.~Dolgovskiy, and
S.~Colombo for stimulating discussions, suggestions and help with auxiliary measurements. }

\section*{References}
\bibliography{manuscript_20140408}

\end{document}